\documentclass[twocolumn, prl, showpacs, amsmath, amssymb, superscriptaddress]{revtex4}

\usepackage{graphicx}
\newcommand{\ket}[1]{\mbox{$\left| \left. #1 \right. \right\rangle$}}

\begin{document}

\title{Four-electron Shell Structures and an Interacting Two-electron System in Carbon Nanotube Quantum Dots}

\author{S. Moriyama}
\affiliation{
Advanced Device Laboratory, The Institute of Physical and Chemical Research (RIKEN), 2-1, Hirosawa, Wako, Saitama 351-0198, Japan
}%
\affiliation{
Interdisciplinary Graduate School of Science \& Engineering, Tokyo Institute of Technology, 4259, Nagatsuta-cho, Midori-ku, Yokohama 226-8503, Japan
}%
\author{T. Fuse}
\affiliation{
Advanced Device Laboratory, The Institute of Physical and Chemical Research (RIKEN), 2-1, Hirosawa, Wako, Saitama 351-0198, Japan
}%
\affiliation{
Interdisciplinary Graduate School of Science \& Engineering, Tokyo Institute of Technology, 4259, Nagatsuta-cho, Midori-ku, Yokohama 226-8503, Japan
}%
\author{M. Suzuki}
\affiliation{
Advanced Device Laboratory, The Institute of Physical and Chemical Research (RIKEN), 2-1, Hirosawa, Wako, Saitama 351-0198, Japan
}%
\affiliation{
CREST, Japan Science and Technology (JST), Kawaguchi, Saitama 332-0012, Japan
}%
\author{Y. Aoyagi}
\affiliation{
Interdisciplinary Graduate School of Science \& Engineering, Tokyo Institute of Technology, 4259, Nagatsuta-cho, Midori-ku, Yokohama 226-8503, Japan
}%
\author{K. Ishibashi}
\email{kishiba@riken.jp}
\affiliation{
Advanced Device Laboratory, The Institute of Physical and Chemical Research (RIKEN), 2-1, Hirosawa, Wako, Saitama 351-0198, Japan
}%
\affiliation{
CREST, Japan Science and Technology (JST), Kawaguchi, Saitama 332-0012, Japan
}%

\date{\today}

\begin{abstract}
Low-temperature transport measurements have been carried out on single-wall carbon nanotube quantum dots in a weakly coupled regime in magnetic fields up to 8 Tesla. Four-electron shell filling was observed, and the magnetic field evolution of each Coulomb peak was investigated, in which magnetic field induced spin flip and resulting spin polarization were observed. Excitation spectroscopy measurements have revealed Zeeman splitting of single particle states for one electron in the shell, and demonstrated singlet and triplet states with direct observation of the exchange splitting at zero-magnetic field for two electrons in the shell, the simplest example of the Hund's rule. The latter indicates the direct analogy to an artificial He atom.
\end{abstract}

\maketitle


Thanks to recent developments in the growth techniques of high quality single-wall carbon nanotubes (SWNTs), 
individual SWNTs displaying quantum dot (QD) behavior have been produced. It is possible that this behavior is clearer than that in the semiconductor QDs~\cite{1}, in terms of the analogy with natural atoms.  Although the experiments on nanotube quantum dots reported so far have revealed various interesting physics, such as shell filling~\cite{2}, Zeeman splitting~\cite{3} and the Kondo 
effect~\cite{4}, they have been observed in various systems with different coupling regimes and different nanotube types, i.e. SWNTs~\cite{2,3,4,5,6} and multi-wall nanotubes (MWNTs)~\cite{7}. In this respect, the physics of nanotube quantum dots does not appear to be systematically understood.

One of the unique features of SWNT QDs is the large zero-dimensional (0-D) energy spacing 
($\Delta$)~\cite{8}, compared with the on-site Coulomb interaction energy ($\delta U$) and the exchange interaction energy ($J$). Besides, $\Delta$ can be as large as the single electron charging energy 
($E_{C} = e^2 / C_{\Sigma}: C_{\Sigma}$ is the self capacitance of the dot). These facts make it possible to  observe shell structures, even though a number of electrons are contained in the dot. Another unique feature is the magnetic field ($B$- field) effect on the single particle state in SWNT QDs, where  Zeeman effect is the only important effect because of the small diameter of SWNTs. These features are in striking contrast to those of standard GaAs/AlGaAs two-dimensional electron gas (2DEG) QDs of submicron size, where the 0-D levels are very likely to be mixed by electron-electron interactions, so that the shell structure can be  observed only in a few electron QDs~\cite{9}, and not in many-electron QDs~\cite{1}. The orbital effect of the $B$-field on 2DEG QD cannot be ignored, which also makes the shell structures much more complicated~\cite{10}. 

\begin{figure}
\includegraphics[width=8cm]{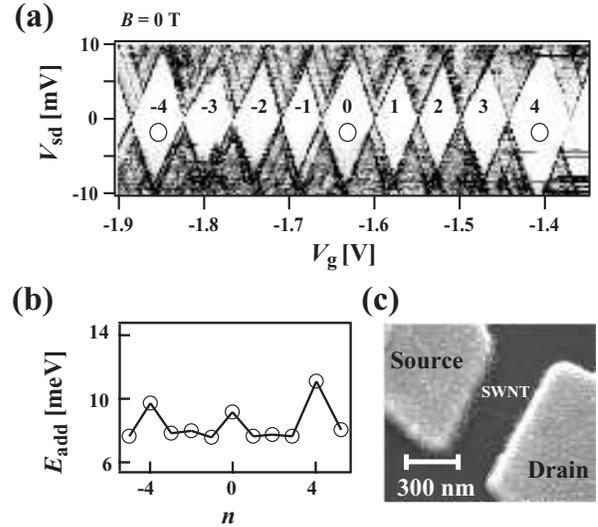}%
\caption{\label{fig1} (a) Gray scale plot of the differential conductance, $dI_{\text{d}}/dV_{\text{sd}}$, as a function of $V_{\text{sd}}$ and $V_{\text{g}}$ at $B = 0$ T. The number, $n$, indicates the number of extra electrons, counted from the diamond around $V_{\text{g}} \sim -1.63 V$. (b) Addition energy ($E_{\text{add}}$) as a function of $n$, determined by the size of the Coulomb diamonds in Fig.1(a). (c) Scanning electron micrograph of the sample.}
\end{figure}

In our experiment, we show that the SWNT QD is suitable for investigating the analogy of the QD with natural atoms, by presenting systematic low-temperature transport data of the closed QDs in magnetic field. The two-and four-electron periodicities have been observed in Coulomb diamonds, but, here, we focus on the latter regime. The excitation spectroscopy revealed the simple Zeeman splitting of single particle states for one electron in the shell. The highlight of the paper is that we, for the first time, have observed an artificial He atom-like behavior for two electrons in the shell, where the textbook model of the interacting two-electron system can be directly applied with observable single and triplet states that have an exchange energy difference at zero magnetic field, the simplest example of the Hund's rule.

A single quantum dot is easily formed in an individual SWNT, just by depositing metallic contacts on it,  which in our case are Ti (Fig.1(c))~\cite{11}. In our fabrication process, a whole nanotube between the two contacts is likely to form a single quantum dot~\cite{12}. All measurements were carried out in a dilution refrigerator at a base temperature of $T_{\text{mix}} = 40$ mK~\cite{13}. A magnetic field ($B$) of up to 8 T was  applied perpendicular to the tube axis. 
Figure 1 shows the Coulomb diamonds with a four-electron periodicity (a), as well as the addition energy with the same periodicity (b), both of which are understood by the four-electron shell model based on the twofold band degeneracy ($A$ and $B$) in addition to the twofold spin degeneracy~\cite{15}.

\begin{figure}
\includegraphics[width=8cm]{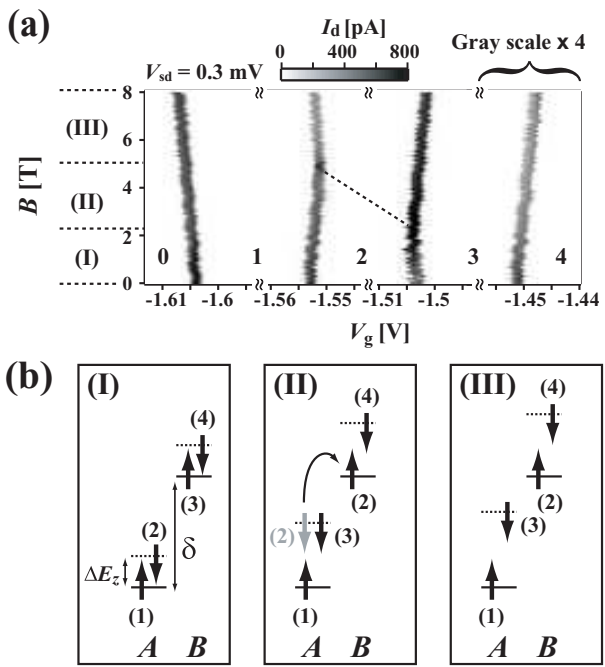}%
\caption{\label{fig2} (a) Magnetic field evolution of Coulomb peaks up to 8 T in the numbered range in Fig. 1(a). $V_{\text{sd}} = 0.3$ mV. The magnetic field range is divided into three parts, depending on the shell filling scheme. (b) Shell filling scheme estimated from the direction of the peak evolution in the three different magnetic field ranges. Single particle states are Zeeman splitted double states with opposite spins. Each number indicates (1) the first, (2) second, (3) third, and (4) fourth electrons which come successively into the shell~\cite{17}. Note that the ``internal spin flip'' occurs in this range.}
\end{figure}

The magnetic field evolution of each Coulomb peak in one period is shown in Fig.2(a), where the current magnitude is also indicated by the gray scale. The $B$ -field can be divided into three ranges, depending on the shell filling scheme. In the low $B$ -field (I) region, each peak shifts linearly in alternate directions, indicating that electrons occupy successive levels from the lowest level, so that the total spin changes between 0 and 1/2 as $n$ is increased, producing an even-odd effect~\cite{16}. However, in the high $B$ -field (III) region, two peaks move in together in the same direction, suggesting spin  polarization. In this case, the total spin changes from $0 \to  1/2 \to 1 \to 1/2 \to 0$ as $n$ increases. The intermediate $B$-field region (II) is between the two kinks that appear in the two lines in the middle. The different kink positions in the two lines suggests that an ``internal 
spin flip'' occurs during the gate sweep, as modeled in Fig.2(b). At lower gate voltages, the second electron occupies 
the $A \downarrow$ state, however, as the gate voltage becomes larger, it flips to the $B \uparrow$ state, so that the third electron can occupy the $A \downarrow$ state. This effect may occur when the energy mismatch ($\delta$) between the $A$ and $B$ state has a $V_{g}$ dependence~\cite{5}, so that, the relative distance between the $A \downarrow$ and $B \uparrow$ state gets closer as 
$V_{\text{g}}$ is swept.

\begin{figure}
\includegraphics[width=8cm]{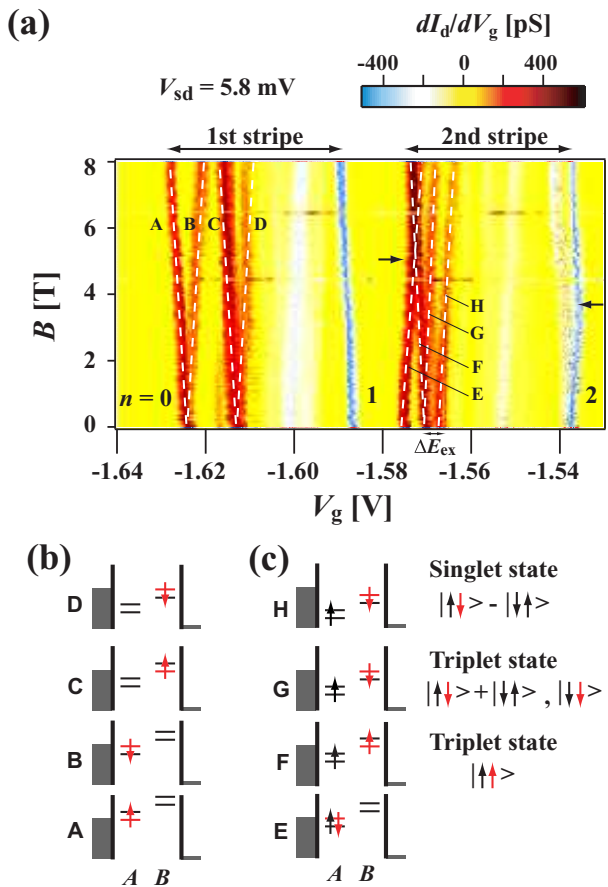}%
\caption{\label{fig3} (color). (a): Excitation spectroscopy measurements in the one- and two-electron systems in a magnetic field in the four-electron shell filling regime. $dI_{\text{d}}/dV_{\text{g}}$ is calculated from $I_{\text{d}}-V_{\text{g}}$ data with $V_{\text{sd}} = 5.8$ mV, and is plotted on a color scale as a function of $V_{\text{g}}$ and $B$.
 Each line from A to H is due to a state shown by the energy diagrams in (b) for the one-electron system and (c) for the two-electron system.}
\end{figure}

The magnetic field evolution of the excited states as well as the ground state can be directly observed in the excitation spectroscopy measurements, and the results are shown in Fig.3(a) for the first ($n = 0 \leftrightarrow 1$) and second ($n = 1 \leftrightarrow 2$) stripes obtained from the Coulomb peaks with large $V_{\text{sd}}$. In the figure, $dI_{\text{d}}/dV_{\text{g}}$ is plotted on a color scale as a function of $V_{\text{g}}$ and $B$. A number of electrons are contained in the dot, however, we can focus only on a single shell composed of four states with similar energies, because the other shells are closed and separated by the large $\Delta$.

The basic idea of the excitation spectroscopy is as follows. Suppose the gate voltage is swept such that the number of electrons in the dot is increased one by one. The current increases whenever a new state comes into the transport window stripe by $V_{\text{sd}}$ because the number of transport channels increases. Once the current has increased to some certain value, it drops to zero (Coulomb blockade) when the first state that  already exists in the transport window comes out of it, resulting in an increment of one electron in the dot~\cite{19}. The red lines indicate positive values, which is an indication of a new state coming into the transport window~\cite{20}. The blue line, which is negative, indicates the sudden drop of the current to zero due to the Coulomb blockade.

In the first stripe in Fig.3(a), the simple $B$-field evolution of each state is observed as lines indicated by A-D. Each line corresponds to Zeeman levels with up and down spins that successively come into the transport window as $V_{\text{g}}$ is increased (Fig.3(b)). The $B$-field dependence of the Zeeman splitting, lines A and B for example, gives a $g$-factor of $1.99 \pm 0.07$, a value similar to that of graphite and a value reported previously~\cite{3,5,21}.

The second stripe, are more interesting in terms of the direct investigation of the interacting two-electron system. An extra-electron is already contained in the dot before the new state comes into the transport window. Each line can basically be understood in a similar way to the case for the one-electron states. Each of the experimentally observed red lines correspond to a measurement of the state which is about to come into the transport window. Equivalently, the measurement corresponds to a projection  of the state. The basic model explaining each line is shown in Fig.3(c). Line F is due to one of the triplet states ($\ket{\uparrow \uparrow}$). (The notation $\ket{\uparrow \downarrow}$, for example, indicates an up-spin in the $A$-subband and a down spin in the $B$-subband). 
The $\ket{\uparrow \downarrow} + \ket{\downarrow \uparrow}$ and $\ket{\downarrow \downarrow}$ states are not possible in this case, because they have higher energy than the $\ket{\uparrow \uparrow}$ state in a $B$-field. Line G occurs due to the triplet states, expressed by $\ket{\uparrow \downarrow} + \ket{\downarrow \uparrow}$ or $\ket{\downarrow \downarrow}$, which are now energetically possible after a slight increase of $V_{\text{g}}$ from the situation for line F~\cite{22}. Of the superposition states, $\ket{\uparrow \downarrow}$ is always detected because the $B \downarrow$ state is used for the measurement. The $\ket{\uparrow \downarrow} + \ket{\downarrow \uparrow}$ and $\ket{\downarrow \downarrow}$ states, which should have a different energy in the $B$-field, are not be able to be distinguished in the present measurement scheme where the on set level or projected state also shifts as a function of the $B$-field. Two states of the triplet are now available for current flow (line H), as compared with one state available for line F. Lines F and G meet at the same $V_{\text{g}}$ position when the $B$-field value goes to zero, which indicates degeneracy of the triplet state ($\ket{\uparrow \uparrow}$, $\ket{\uparrow \downarrow} + \ket{\downarrow \uparrow}$ and $\ket{\downarrow \downarrow}$ ) at $B = 0$ T. One might think three lines should be observed, associated with the triplet state. However, due to the above mentioned measurement scheme, two lines can be observed. 
Line H, which runs just next to line G, is attributed to the singlet state, $\ket{\uparrow \downarrow} - \ket{\downarrow \uparrow}$, with a finite energy larger than the energy of the triplet state. The separation ($\Delta E_{\text{ex}}$) between lines F and H at $B = 0$ T directly corresponds to the energy difference between the singlet and triplet states, the exchange energy $J$. This is a direct demonstration of the simplest example of the Hund's rule, in the sense that the higher spin state, $S = 1$ in the present case, is likely to occur due to the exchange effect which lowers the total energy. 

We may also show the excitation spectroscopy data for the three- and four-electron shell filling regimes. However, the overall signals are rather small, compared with those in the one- and two-electron regimes.  Simple Zeeman splitting can be observed when the filled state comes out of the transport window (current decreasing regime), but there are features that are not fully understood. We will report on this regime at our next opportunity with a more convincing interpretation.

It should be noted that the lines E and F cross in the second stripe, as indicated by the arrow, while  crossing is not observed in the first stripe. The line crossing observed in the second stripe, is closely related to the kink observed in the blue line, since the blue line should be a replica of the leftmost red  line. The blue lines occur when the state that has first come into the transport window comes out of it, and the system is Coulomb blockaded. In fact, the expected behavior is shown in both the leftmost red and blue lines except for the different kink position. This effect, indicated by the arrows, is again explained by the $V_{\text{g}}$ dependent $\delta$, as is the case in Fig.2(a). Actually the slope of the line connected by the two arrows is consistent with that of the dotted line in Fig.2(a).

Having understood the qualitative behavior of shell filling and the two-electron interaction behavior in the SWNT quantum dot, we now estimate various energy scales associated with the dot. The addition energies for each Coulomb diamond that shows the four-electron periodicity contain information on interaction energies as well as the single particle level spacing~\cite{18}.
Based on the Hamiltonian given in Ref.~\cite{18}, the energy values are obtained as 
$\delta = 1.7 - 0.006 \Delta V_{\text{g}}$ meV, $\Delta = 5.9$ meV, 
$E_{C} = 6.7$ meV, $\delta U = 0.4$ meV, $J = 0.5$ meV. $\Delta V_{\text{g}}$ is measured from the 1st Coulomb peak position ($n = 0 \leftrightarrow 1$) at $B = 0$. $J$ and $\delta$ at $\Delta V_{\text{g}} = 0$ were obtained directly from the exchange splitting ($\Delta E_{\text{ex}}$) in Fig.3(a) and the first excited line in the Coulomb diamond of Fig.1(a), respectively. 
The condition, $\delta  < \Delta /2$, necessary for observation of the four-electron periodicity, is, in fact, satisfied. $\Delta$ as large as $E_{C}$ is unique for SWNT QD. The simple theoretical estimate of $\Delta (= 5.6$ meV), based on $hv_{F}/ 2 L$ 
($L$, the length of the contact gap, is 300 nm and equivalent to the dot size, $v_{F} = 8.1 \times 10^{5}$ m/s.) where subband degeneracy is assumed, is in good agreement with that obtained $\Delta$ ($= 5.9$ meV) in the experiment. This fact indicates that the quantum levels indeed originate from one-dimensional confinement of electrons in the tube-axis direction.
The estimated energy parameters normalized by $\Delta$ appear to be consistent with the previously reported~\cite{6} and predicted~\cite{18} values, and confirm the unique condition in the SWNT QD, which is mentioned in the introductory part. It is interesting to note that the on-site Coulomb energy and the exchange interaction energies are three or four orders smaller in the SWNT QD than those values of the natural He atom \cite{24}, which might be reasonable because of the large difference in the space where electrons are confined. 

In summary, we have carried out low temperature transport measurements in individual SWNT quantum dots. The four-electron shell filling regime has been carefully investigated, and the magnetic field evolution of each Coulomb peak has revealed the different shell fillings in low, high and intermediate magnetic field ranges.  Excitation spectroscopy measurements have been carried out in the one and two-electron regimes, and the interacting two-electron model in a magnetic field was directly observed. 

We thank Prof. M. Eto of Keio University for useful discussions and suggestions. We also enjoyed discussions with Dr. A. Furusaki and Dr. Y. Ishiwata of RIKEN, and Prof. C.J.P.M. Harmans of Delft University of Technology. We also appreciate continuous encouragements from Prof. K. Gamo and Prof. T Sugano.


\begin{thebibliography}{99}
\bibitem[{8()}]{8}
\bibinfo{note}{Throughout the paper, we distinguish the 0-D energy spacing, $\Delta$, and the subband  mismatch, $\delta$. $\Delta$ is determined by the one-dimensional confinement in the tube axis direction and is the same for both subbands, while $\delta$ is determined by the mismatch between the two ($A$ and $B$) subbands which originates from the unique band structure of the nanotube at the Fermi energy.}
\bibitem[{13()}]{13}
\bibinfo{note}{The effective temperature, $T_{e} = 240$ mK, was estimated by fitting the Coulomb peak measured at 
$V_{\text{sd}} = 10 \mu$ V with the quantum version of the theoritical expression of the peak~\cite{14}.}
\bibitem[{17()}]{17}
\bibinfo{note}{The additional energies, $E_{\text{add}}(n)$ of each region are as follows.
For $J + \delta U + \Delta E_{z} < \delta$ (low magnetic field):
$E_{\text{add}}(1) = \Delta + E_{C} - \delta U - \delta - \Delta E_{z}$, 
$E_{\text{add}}(2) = E_{\text{add}}(4) = E_{C} + \delta U +J + \Delta E_{z}$, 
$E_{\text{add}}(3) = E_{C} + \delta - \delta U - \Delta E_{z}$.
For $J + \delta U + \Delta E_{z} > \delta$ (high magnetic field):
$E_{\text{add}}(1) = \Delta + E_{C} - \delta U - \delta - \Delta E_{z}$, 
$E_{\text{add}}(2) = E_{\text{add}}(4) = E_{C} + \delta$, 
$E_{\text{add}}(3) = E_{C} + \delta U + 2 J - \delta + \Delta E_{z}$. 
These equations are deduced by calculating the Hamiltonian of the nanotube quantum dot, given in 
Ref.~\cite{18}.}
\bibitem[{20()}]{20}
\bibinfo{note}{Precisely speaking, it is likely that some of the states in the lower closed shell may also exist in the transport window when the Coulomb blockade is lifted, since the applied $V_{\text{sd}} = 5.8$ mV is very close to the estimated $\Delta$. Some of the white lines, such as those which appear at $V_{\text{g}} \sim -1.6$ V and $-1.555$ V in Fig.3 (a) are related to this fact. However, the effect of the lower closed shell on the electronic state of the open shell can be ignored.}
\bibitem[{21()}]{21}
\bibinfo{note}{There appears to be a slight deviation form the straight line above $B = 4$ T for the line D. This may be explained by considering that the magnetic field evolution of the onstripe of the new state coming into the transport window may be strongly affected by the magnetic field dependence of other already existing other states.}
\bibitem[{22()}]{22}
\bibinfo{note}{The $\ket{\downarrow \downarrow}$ is possible in such a case where the original 
$A \uparrow$ electron is replaced by a $A \downarrow$ electron before the $B \downarrow$ state comes into the transport window. Otherwise, the spin blockade mechanism does not allow the $\ket{\downarrow \downarrow}$ configuration~\cite{23}. 
(The absolute change of $S_{z}$ becomes more than 1/2.)}
\end{thebibliography}
\end{document}